\documentclass[aps,preprintnumbers,nofootinbib,superscriptaddress,11pt]{revtex4}
\usepackage[hypertex]{hyperref}
\usepackage[dvipdfmx]{graphicx} 
\usepackage{amsmath,amssymb,amsfonts,bm,cancel} 

               \def\n{\nu}      \def\s{\sigma}         

\def\dg{\dagger}

    \newcommand{\To}{\Rightarrow}

\renewcommand{\Re}{{\rm Re \, }} \renewcommand{\Im}{{\rm Im \, }}

\newcommand{\row}[2]{ \begin{pmatrix}  #1 & #2   \end{pmatrix}  }
\newcommand{\column}[2]{ \begin{pmatrix}  #1 \\ #2 \\  \end{pmatrix} }

\newcommand{\Column}[3]{ \begin{pmatrix} #1 \\ #2 \\ #3 \end{pmatrix} }
\newcommand{\diag}[2]{ \begin{pmatrix}  #1 & 0 \\ 0 & #2 \\   \end{pmatrix}  }
\newcommand{\offdiag}[2]{ \begin{pmatrix} 0 & #1 \\ #2 & 0 \\   \end{pmatrix} }

\begin{document}


\title{\large \bf A formula by $LDL^{T}$ decomposition for the minimal type-I seesaw mechanism \\ and conditions of $CP$ symmetry in an arbitrary basis}

\preprint{STUPP-21-253}
\author{Masaki J. S. Yang}
\email{yang@krishna.th.phy.saitama-u.ac.jp}
\affiliation{Department of Physics, Saitama University, 
Shimo-okubo, Sakura-ku, Saitama, 338-8570, Japan}


\begin{abstract} 

In this paper, defining a formula by $LDL^{T}$ decomposition for the minimal type-I seesaw mechanism, we obtain conditions of $CP$ symmetry for the neutrino mass matrix $m$ in an arbitrary basis.  
The conditions are found to be ${\rm Re\,} (M_{22} a_{i} - M_{12} b_{i}) \, {\rm Im \,} ( M_{22} a_{j} - M_{12} b_{j}) = - \det M \, {\rm Re\,} b_{i} \, {\rm Im \,} b_{j}$ or $ = - \det M \, {\rm Im \,} b_{i} \, {\rm Re\,} b_{j}$ for the Yukawa matrix $Y_{ij} = (a_{j}, b_{j})$ and the right-handed neutrino mass matrix $M_{ij}$. 
In other words, the real or imaginary part of $b_{i}$ must be proportional to the real or imaginary part of the quantity $(M_{22} a_{i} - M_{12} b_{i})$.

\end{abstract} 

\maketitle

\section{Introduction}

$CP$ violation (CPV) in the neutrino oscillation has been strongly suggested by T2K \cite{T2K:2021xwb} and NO$\n$A \cite{NOvA:2021nfi}. 
For this reason, CPV in the lepton sector has been widely studied using the seesaw mechanism \cite{Minkowski:1977sc, GellMann:1980v, Yanagida:1979as}.
In the analysis of seesaw relations, it is common to reduce parameters by diagonalization and/or phase redefinition \cite{Casas:2001sr,Ibarra:2003up, Barger:2003gt}. 
However, such a representation obscures symmetries and relationships of the original Lagrangian.  
Information of $CP$ phases is often lost in such parameterization, even if many of phases are non-physical.
It would be somewhat important to investigate  structures of $CP$ phases in the Lagrangian without redefinitions. 
In this paper, we analyze conditions of $CP$ symmetry for the light neutrino mass matrix $m$ 
 defining a formula  by $LDL^{T}$ decomposition \cite{Yang:2021arl} for the minimal type-I seesaw mechanism 
\cite{Ma:1998zg, King:1998jw, Frampton:2002qc, Xing:2020ald, Guo:2003cc, Mei:2003gn, Chang:2004wy, Guo:2006qa, Kitabayashi:2007bs, He:2009pt, Yang:2011fh, Harigaya:2012bw, Kitabayashi:2016zec, Bambhaniya:2016rbb, Li:2017zmk, Liu:2017frs, Shimizu:2017fgu, Shimizu:2017vwi, Nath:2018hjx, Barreiros:2018bju, Nath:2018xih,  Wang:2019ovr, Zhao:2020bzx}. 
As a result, conditions are obtained for Lagrangian parameters in an arbitrary basis.
Furthermore, we discuss relationships between the obtained solution and generalized $CP$ symmetry (GCP) 
\cite{Ecker:1981wv, Ecker:1983hz, Gronau:1985sp, Ecker:1987qp,Neufeld:1987wa,Ferreira:2009wh,Feruglio:2012cw,Holthausen:2012dk,Ding:2013bpa,Girardi:2013sza,Nishi:2013jqa,Ding:2013hpa,Feruglio:2013hia,Chen:2014wxa,Ding:2014ora,Ding:2014hva,Chen:2014tpa,Chen:2015siy,Li:2015jxa,Turner:2015uta, Rodejohann:2017lre, Penedo:2017vtf,Nath:2018fvw, Yang:2020qsa, Yang:2020goc, Yang:2021smh, Yang:2021xob}.

This paper is organized as follows. 
The next section gives a formula for the minimal type-I seesaw mechanism. 
In Sec.~3, we discuss conditions for $CP$ symmetry in an arbitrary basis. 
In Sec.~4, some relationships between the $CP$-invariant conditions and GCP are discussed. 
The final section is devoted to a summary.

\section{A formula for the minimal type-I seesaw mechanism}

In the beginning, we define a formula by $LDL^{T}$ decomposition \cite{Yang:2021arl} 
for the minimal type-I seesaw mechanism \cite{Ma:1998zg, King:1998jw, Frampton:2002qc, Xing:2020ald}. 
By setting the vacuum expectation value of Higgs to one, 
the two mass matrices of neutrinos are defined as follows, 
\begin{align}
Y = 
\begin{pmatrix}
a_{1} & b_{1} \\
a_{2} & b_{2} \\
a_{3} & b_{3} \\
\end{pmatrix} \equiv 
 \Column{~~ \bm Y_{1} ~~}{~~ \bm Y_{2} ~~}{~~ \bm Y_{3} ~~} \, , 
~~~ 
M =
\begin{pmatrix}
M_{11} & M_{12} \\ M_{21} & M_{22}
\end{pmatrix} \equiv \column{~~ \bm M_{1} ~~}{~~ \bm M_{2} ~~} \, . 
\label{1}
\end{align}
Here, two-dimensional complex row vectors
 $(\bm Y_{i})_{j} \equiv (Y_{ij})$ and $(\bm M_{i})_{j} \equiv M_{ij}$ have mass dimension one.
Let us consider a case where $M$ and its eigenvalues $M_{i}$ are hierarchical; 
\begin{align}
|M_{22}| \gg |M_{12}| \, , |M_{11}| \, ,  ~~ 
M_{2} \simeq M_{22} + {M_{12}^{2} \over M_{22}} \, ,  ~~ 
M_{1} \simeq {\det M \over M_{2}} \, . 
\end{align}
As in the case of the type-I seesaw mechanism, we perform an approximate spectral decomposition for $M^{-1}$; 
\begin{align}
M^{-1} =  {1\over \det M} 
\begin{pmatrix}
M_{22} & - M_{12} \\ - M_{12} & M_{11}
\end{pmatrix}
 & = 
{1\over \det M} 
\begin{pmatrix}
M_{22} & - M_{12} \\ - M_{12} & {M_{12}^{2} \over M_{22}}
\end{pmatrix}
+ \diag{0}{ 1/ M_{22}} \,  \\
 & = M^{(1)} + M^{(2)} \, .
\label{spectal}
\end{align}
The eigenvalues\footnote{The modulus of eigenvalues correspond to
 singular values of the $M^{(1,2)}$, $(|M_{22}|^{2} + |M_{12}|^{2}) / | M_{22} \det M|$ and $|M_{22}|^{-1}$ in the limit of $M_{12}/ M_{22} \to 0$.} of the $M^{(1,2)}$ are $(M_{22}^{2}+ M_{12}^{2}) / M_{22} \det M$ and $M_{22}^{-1}$, and it corresponds to the first-order perturbation of the spectral decomposition. 

The mass matrix of the light neutrinos $m$ is found to be 
\begin{align}
m 
&= Y (M^{(1)} + M^{(2)}) Y^{T} \equiv m^{(1)} + m^{(2)} \\
& = 
{ M_{22} \over \det M}
\begin{pmatrix}
\tilde a_1^2 & \tilde a_1 \tilde a_2 & \tilde a_1 \tilde a_3 \\
\tilde a_1 \tilde a_2 & \tilde a_2^2 & \tilde a_2 \tilde a_3 \\
\tilde a_1 \tilde a_3 & \tilde a_2 \tilde a_3 & \tilde a_3^2 \\
\end{pmatrix}
+ {1\over M_{22}}
\begin{pmatrix}
b_{1}^{2} & b_{1} b_{2} & b_{1} b_{3} \\
b_{1} b_{2} & b_{2}^{2} & b_{2} b_{3} \\
b_{1} b_{3} & b_{2} b_{3} & b_{3}^{2}
\end{pmatrix} \, , 
\label{formula}
\end{align}
where
\begin{align}
\tilde a_{i} \equiv {\det (\bm Y_{i} , \, \bm M_{2}) \over M_{22}}
= a_{i} - b_{i} {M_{12} \over M_{22}} \, . 
\label{7}
\end{align}
This is equivalent to a formula for the type-I seesaw mechanism by setting $M_{33} \to \infty$.
Since no approximation is used obviously, this formula is valid in any basis. 
Therefore, it can be useful for various analyses, such as flavor symmetries, GCPs, and fine-tunings of the seesaw mechanism.

For a unitary matrix $U$ diagonalizing $m$ as $U^{T} m U = m^{\rm diag}$, 
the values of neutrino masses are expressed without radical symbols $(\sqrt{})$; 
\begin{align}
m_{i} = {M_{22} \over |M|} (\tilde {\bm a} \cdot \bm u_{i})^{2} + {1\over M_{22}} (\bm b \cdot \bm u_{i}) \, . 
\end{align}
Here, $|M| \equiv \det M$, $\tilde {\bm a}_{i} \equiv \tilde a_{i}$, $\bm  b_{i} \equiv b_{i}$, $(\bm u_{i})_{j} \equiv U_{ji}$ are three-dimensional vectors, and 
$(\bm u \cdot \bm v) \equiv \sum_{i = 1}^{3} u_{i} v_{i}$ is the inner product without Hermitian conjugation.

A deformed Yukawa matrix $\tilde Y$ is defined by 
\begin{align}
\tilde Y \equiv 
\begin{pmatrix}
 &  \\[-8pt] \tilde {\bm a} \, & \bm b\\[-8pt]  &  \\
\end{pmatrix} \, 
\equiv (\bm a - \bm b {M_{12} \over M_{22}}  \, , \, \bm b) \, 
= Y 
\begin{pmatrix}
1 & 0 \\ - {M_{12} \over M_{22}} & 1
\end{pmatrix}
\equiv Y L \, , 
%
\end{align}
where $L$ is a lower unitriangular matrix 
\begin{align}
L = 
\begin{pmatrix}
1 & 0 \\ - {M_{12} \over M_{22}} & 1
\end{pmatrix} \, , 
~~~ 
L^{-1} =
\begin{pmatrix}
1 & 0 \\  {M_{12} \over M_{22}} & 1
\end{pmatrix} \, ,
\end{align}
that has all the diagonal entries equal to one. 
The mass matrix of heavy neutrinos $M$ is diagonalized by $L$ as
\begin{align}
\tilde M^{-1} = L^{-1} M^{-1} (L^{-1})^{T}
= \diag{{M_{22} \over \det M} }{1\over M_{22}} \, , 
\label{tildeM}
\end{align}
and $m = \tilde Y \tilde M^{-1} \tilde Y^{T}$ holds. 
This is called the $LDL^{T}$ (or generalized Cholesky) decomposition of a symmetric matrix. 

This formula can be regarded as an extension of the natural representation \cite{Barger:2003gt}.
Since it is possible to reverse the diagonalization by $L$ using only $M_{12} / M_{22}$, 
one advantage is that the information of Lagrangian includes $CP$ phases is treated 
without solving quadratic equations. 

\section{Conditions for $CP$ symmetry in an arbitrary basis}

In this section, using the formula~(\ref{formula}), we investigate conditions for $CP$ symmetry for the minimal type-I seesaw mechanism.
First, Eq.~(\ref{formula}) is rewritten as 
\begin{align}
m = {M_{22} \over |M|} \tilde {\bm a} \otimes \tilde {\bm a}^{T} + {1\over M_{22} } \bm b \otimes \bm b^{T} \, . 
\label{11}
\end{align}
By redefining phases of the right-handed neutrinos, we can choose a basis in which $M_{22}$ and $|M|$ are real-positive. 
It is easy to incorporate these overall phases into the final result by inverse  redefinitions.

If the matrix $m$ satisfies the $CP$ symmetry, 
each imaginary part must cancel in Eq.~(\ref{11}); 
\begin{align}
\Im m_{ij} = 
{M_{22} \over |M|} \Im (\tilde a_{i} \tilde  a_{j}) &+ {1\over M_{22} }\Im (b_{i} b_{j}) = 0 \, .
\end{align}
By separating the imaginary parts of products, 
\begin{align}
 {M_{22} \over |M|} (\Re \tilde a_{i} \, \Im \tilde a_{j} + \Im \tilde a_{i} \, \Re \tilde a_{j} )
+ {1\over M_{22} } (\Re  b_{i} \, \Im  b_{j} +  \Im  b_{i} \, \Re  b_{j}) = 0 \, . 
\label{cond}
\end{align}
Furthermore, 
by considering cross products with $\Im b_{i}$ and $\Im b_{j}$ for $i$ and $j$ components, 
the term containing $1/M_{22}$ becomes zero;
\begin{align}
 {M_{22} \over |M|} [(\Im \bm b \times \Re \tilde {\bm a} ) \otimes (\Im \tilde {\bm a} \times \Im {\bm b})^{T} 
 + (\Im \bm b \times \Im \tilde {\bm a}) \otimes (\Re \tilde {\bm a} \times \Im \bm b)^{T}] = \bm 0 \otimes \bm 0 \, . 
\end{align}
This is an antisymmetric condition for a matrix. 
However, since the only antisymmetric matrix with a rank less than one is the zero matrix, we obtain 
\begin{align}
\Im \bm b \times \Re \tilde {\bm a} = \bm 0 ~~ {\rm or} ~~ \Im \tilde {\bm a} \times \Im {\bm b} = \bm 0 \, , 
~~~
 \Re \tilde {\bm a} \propto \Im \bm b ~~ {\rm or} ~~  \Im \tilde {\bm a} \propto \Im \bm b \, .
\end{align}
Similarly, $\Re \tilde {\bm a}\propto \Re \bm b$ or $\Im \tilde {\bm a}\propto \Re \bm b$ can be shown. 
As a result, the conditions that $m$ is $CP$-invariant are equivalent to 
\begin{align}
M_{22}^{2} \, \Re \tilde a_{i} \, \Im \tilde a_{j} = 
-  |M| \, \Re  b_{i} \, \Im  b_{j} ~~ {\rm or} ~~ 
-  |M| \, \Im  b_{i} \, \Re  b_{j} \, .
\label{cond3}
\end{align}
Thus, the real and imaginary parts of $\tilde a_{i}$ are proportional to those of $b_{i}$, and their coefficients are determined by Eq.~(\ref{cond3}).
At first glance, these conditions appear to give nine constraints.
However, since a matrix $A \equiv \Re \tilde a_{i} \, \Im \tilde a_{j}$ with rank one satisfies $A_{ii} A_{jj} = A_{ij} A_{ji}$, it is necessary and sufficient to give $A_{1i}$ and $A_{j1}$. Therefore, there are only five independent constraints.

By the conditions~(\ref{cond3}), we can express $\Re \tilde {\bm a}$ and $\Im \tilde {\bm a}$ 
for given $\Re \bm b$ and $\Im \bm b$;  
\begin{align}
(M_{22} \, \Re \tilde {\bm a} \, , M_{22} \, \Im \tilde {\bm a}) = ( r \sqrt{|M|}\, \Re \bm b \, , - {1\over r} \sqrt{|M|} \, \Im \bm b) 
~~ {\rm or} ~~  ( r' \sqrt{|M|} \, \Im \bm b \, , - {1\over r'} \sqrt{|M|}  \, \Re \bm b) \, .
\label{solution}
\end{align}
Here,  $r$ and $r'$ are real constants defined by non-zero $\Im \tilde a_{i}$ and $\Re b_{j}$ or $\Im b_{j}$, 
\begin{align}
r \equiv  {M_{22} \over \sqrt{|M|}} {\Re \tilde a_{j} \over \Re b_{j}} \, 
= - {\sqrt{|M|} \over M_{22} } {\Im b_{i} \over \Im \tilde a_{i}} ,  ~~~ 
r' \equiv  {M_{22} \over \sqrt{|M|}} {\Re \tilde a_{i} \over \Im b_{i}}
= - {\sqrt{|M|} \over M_{22} } {\Re b_{j} \over \Im \tilde a_{j}}  \, .
\end{align}
%
The other solution corresponds to $- i \tilde {\bm a}^{*}$ for a given solution $\tilde {\bm a}$.

$\Re \tilde {\bm a} =\bm 0$ or $\Im \tilde {\bm a} =\bm 0$ are special cases  
where the denominators of $r^{(\prime)}$ or $1/r^{(\prime)}$ cannot be defined 
and the magnitudes of $\tilde {a_{i}}$ and $b_{i}$ are unconstrained. 
In such cases, the elements of $\tilde Y$ are real or pure imaginal;
\begin{align}
 \tilde Y &= (\tilde a_{i} , b_{i}) = 
\{ ( x_{i} , y_{i}) \, , ~ (  x_{i} , i y_{i}) \, , ~ 
 ( i x_{i} ,  y_{i}) \, , ~ ( i  x_{i} ,  i y_{i}) \} \, ,
\end{align}
with real parameters $x_{i} , y_{i} \in \mathbb R$. 
In each case, $\tilde Y$ has a (generalized) $CP$ symmetry; 
\begin{align}
\tilde Y^{*} = \{ \tilde Y \, , ~ \tilde Y \s_{3} \, , ~ - \tilde Y \s_{3} \, , ~ - \tilde Y \} \, , 
\label{trivial}
\end{align}
where $\s_{3} \equiv {\rm diag} (1, -1) \, .$

For example, if $\tilde Y$ has the canonical $CP$ and $\Im \tilde a_{i} = \Im b_{i} = 0$ holds,
the relation $a_{i} = \tilde a_{i} + {M_{12} \over M_{22}} b_{i}$ leads to 
\begin{align}
& \Re a_{i} =  \Re \tilde a_{i} + \Re {M_{12} \over M_{22}} \Re b_{i} \, , \\
& \Im a_{i} =  \Im {M_{12} \over M_{22}} \Re b_{i} \, , 
\label{Ima}
\end{align}
for any $i$. 
In other words, the imaginary part of $\bm a$ and the real part of $\bm b$ are required to be proportional. 
If $\Im M_{12} = 0$, $\Im M = 0$ holds since we chose $\Im M_{22} = \Im |M|=0$ in this basis.
It leads to $\Im a_{i} = 0$ and $\Im Y = 0$, {\it i.e.}, the neutrino sectors $M$ and $Y$ are $CP$-invariant.
However, Eq.~(\ref{Ima}) shows the existence of other nontrivial solutions.
They can be regarded as a kind of alignment conditions that are required for the naturalness of the seesaw mechanism \cite{Yang:2021byq}.
For the remaining three cases, similar alignments holds between 
$\Re a_{i}, \Im a_{i}$ and $\Re b_{i}, \Im b_{i}$.

\subsection{Understanding from the original raw formula}

We can understand the result~(\ref{cond3}) without $LDL^{T}$ decomposition.
By adjusting coefficients in Eq.~(\ref{cond3}) and rewriting $\tilde {\bm a}$ to $\bm a$, 
\begin{align}
 \Re (M_{22} a_{i} - M_{12} b_{i}) \, \Im ( M_{22} a_{j} - M_{12} b_{j}) = 
- |M| \, \Re  b_{i} \, \Im  b_{j} ~~ {\rm or} ~~ 
- |M| \, \Im  b_{i} \, \Re  b_{j} \, .
\label{cond4}
\end{align}
With some transformation, we obtain 
\begin{align}
 \Im [ (a_{i} M_{22} - b_{i} M_{12}) a_{j} ] &= 
 - \Im [  ( b_{i} M_{11}  - a_{i} M_{12}) b_{j}] \, . 
\label{cond5}
\end{align}
This is equivalent to the condition $m = m^{*}$ in the original raw formula.
From Eq.~(\ref{1}), the mass matrix $m$ is written as 
\begin{align}
m &= Y M^{-1} Y^{T} = 
{1 \over |M|} 
\begin{pmatrix}
a_{1} & b_{1} \\
a_{2} & b_{2} \\
a_{3} & b_{3} \\
\end{pmatrix}
\begin{pmatrix}
- (\bm M_{2} \times \bm Y_{1})_{z} & - (\bm M_{2} \times \bm Y_{2})_{z} & -(\bm M_{2} \times \bm Y_{3})_{z} \\
(\bm M_{1} \times \bm Y_{1})_{z} & (\bm M_{1} \times \bm Y_{2})_{z} & (\bm M_{1} \times \bm Y_{3})_{z} \\
\end{pmatrix} \, , \\
m_{ij} & = {1\over |M|} [- a_{i} (M_{21} b_{j} - M_{22} a_{j}) + b_{i}  (M_{11} b_{j} - M_{12} a_{j})] \, , 
\end{align}
where $(\bm M_{i} \times \bm Y_{j})_{z} \equiv M_{i1} Y_{j2} - M_{i2} Y_{j1}$.  
Since the $CP$-invariance requires cancellation of the imaginary parts between the two terms, Eq.~(\ref{cond5}) is obtained as the conditions.
%

\section{Understanding from generalized $CP$ symmetry} 

The solution~(\ref{solution}) can also be understood from GCP.
Although some results in this section are well known, an analysis is performed for confirmation of the conditions.
By defining $X \equiv \tilde Y \sqrt{\tilde M^{-1}}$, $m$ is written only in $X$;
\begin{align}
X = \row{ \sqrt{\dfrac{M_{22}}{|M|}} \tilde {\bm a}}{\sqrt{\dfrac{1}{M_{22}}}\bm b} \, , ~~~
m = X X^{T} \, .
\end{align}
In order for $m$ to have $CP$ symmetry, 
$X = (u , v)$ must satisfy the following GCP with a complex orthogonal matrix $O$; 
\begin{align}
X^{*} O =
%
\pm X = \pm (u \, , \, v) \, .
\label{GCP}
\end{align}

Since $X O^* O = \pm X^* O = X$ holds from a conjugation of Eq.~(\ref{GCP}), 
the matrix $O$ satisfies $O^* = O^{-1}=O^T$ and $O= O^{\dg}$. 
In other words, $O$ is a Hermitian orthogonal matrix.
The situation is divided into two cases with $\det O = \pm 1$
\cite{Chen:2016ptr, Li:2017zmk, Barreiros:2018bju, Zhao:2021dwc}; 
\begin{align}
 O = 
\begin{pmatrix}
 \cosh x & - i \sinh x \\
i \sinh x &  \cosh x \\
\end{pmatrix}
~~ {\rm or } ~~
\begin{pmatrix}
 \cos x & \sin x \\
 \sin x & -\cos x \\
\end{pmatrix} \, , 
\label{O}
\end{align}
where $x$ is a real parameter.

When $O$ is real, the GCP relation is
\begin{align}
X^{*} O = (u^{*} , v^{*}) 
\begin{pmatrix}
\cos x & \sin x \\
\sin x & - \cos x \\
\end{pmatrix}
= ( c_{x} u^{*} +  s_{x} v^{*} , \, s_{x} u^{*} - c_{x} v^{*})
=  \pm (u \, , \, v) = \pm X \, .
\end{align}
where $c_{x} \equiv \cos x , s_{x} \equiv \sin x$.  
From this, $u^{*}$ and $u$ can be solved for $v$ and $v^{*}$;
\begin{align}
 u^{*} = {1 \over s_{x}} (\pm v + c_{x} v^{*}) \, ,  ~~~
 u = \pm {c_{x} \over s_{x}} (\pm v + c_{x} v^{*}) \pm s_{x} v^{*} 
= {c_{x} \over s_{x}} v \pm  {1 \over s_{x}} v^{*} \, .
\end{align}
By rewriting them with the real and imaginary parts of $u$ and $v$,
\begin{align}
 \sqrt{\dfrac{M_{22}}{|M|}} \Re \tilde {\bm a}  = 
\Re u & 
= { c_{x} \pm 1 \over s_{x}} \Re v
= \{\cot {x \over 2} \, , \, -\tan {x \over 2} \} \sqrt{\dfrac{1}{M_{22}}} \Re \bm b \, ,  
\label{Reu2} \\
 \sqrt{\dfrac{M_{22}}{|M|}} \Im \tilde {\bm a}  = 
\Im u & 
= {c_{x} \mp 1 \over s_{x}} \Im v 
= \{-\tan {x \over 2} \, , \, \cot {x \over 2} \} \sqrt{\dfrac{1}{M_{22}}} \Im \bm b  
\label{Imu2} \, .
\end{align}
Obviously, these coefficients satisfy 
\begin{align}
{ c_{x} \pm 1 \over s_{x}} \times {c_{x} \mp 1 \over s_{x}} = -1,
\end{align}
and it corresponds to the first solution of Eq.~(\ref{solution}). 
The signs $\pm$ come from the parity for GCP~(\ref{GCP})  and it interchanges $r$ and $1/r$. 

The other solution with $|O| = 1$ satisfies
\begin{align}
X^{*} O = (u^{*} , v^{*}) 
\begin{pmatrix}
 \cosh x & - i \sinh x \\
i \sinh x & \cosh x \\
\end{pmatrix}
= ( ch_{x} u^{*} + i sh_{x} v^{*} , - i sh_{x} u^{*} + ch_{x} v^{*})
= \pm (u \, , \, v) = \pm X \, , 
\end{align}
where $ch_{x} \equiv \cosh x , sh_{x} \equiv \sinh x$.
A similar calculation yields
\begin{align}
\sqrt{\dfrac{M_{22}}{|M|}} \Re \tilde {\bm a} =
\Re u & 
= {- ch_{x} \mp 1 \over sh_{x}} \Im v 
= {- ch_{x} \mp 1 \over sh_{x}} \sqrt{\dfrac{1}{M_{22}}} \Im \bm b\, , 
\label{Reu} \\
\sqrt{\dfrac{M_{22}}{|M|}} \Im \tilde {\bm a} = 
\Im u & 
= {ch_{x} \mp 1 \over sh_{x}} \Re v 
= {ch_{x} \mp 1 \over sh_{x}} \sqrt{\dfrac{1}{M_{22}}} \Re \bm b  \, .
\label{Imu}
\end{align}
Since the product of the two coefficients becomes unity, 
\begin{align}
{- ch_{x} \mp 1 \over sh_{x}} \times {ch_{x} \mp 1 \over sh_{x}} 
= {- ch_{x}^{2} + 1 \over sh_{x}^{2}} = -1 \, , 
\end{align}
it corresponds to the second solution of Eq.~(\ref{solution}). 

These solutions can also be understood by matrices.
For $|O| = +1$, we can write 
\begin{align}
1 \pm O^{T} & = (1 \pm c_{z} ) 1_{2} \pm i  s_{z} \s_{2} \, ,  
\label{49}
\end{align}
where $\s_{i}$ are the Pauli matrices.
From $\s^{2}_{2} = 1$, a certain relation exists between $1\pm O^T$; 
\begin{align}
(1\pm O^{T}) {1 \mp c_{z} \over s_{z}} = s_{z} 1_{2} \pm i  (1 \mp c_{z}) \s_{2}
= \pm (1 \mp O^{T}) i \s_{2} \, . 
\end{align}
From $X^{*} = \pm X O^{T}$, 
it leads to a relation between the real and imaginary parts of $X$, 
\begin{align}
\Re X = X {1 \pm O^{T} \over 2} 
= \pm {s_{z} \over 1 \mp c_{z} } X {1 \mp O^{T} \over 2} i \s_{2}
= \pm {s_{z} \over 1 \mp c_{z} } i\, \Im X \offdiag{1}{-1}  . 
\end{align}
In other words, 
\begin{align}
\begin{pmatrix}
 &  \\[-8pt] \Re \tilde {\bm a} \, & \Re \bm b\\[-8pt]  &  \\
\end{pmatrix} \diag{\sqrt{\frac{M_{22}}{|M|}} }{\sqrt{\frac{1}{M_{22}}}}
=
  { sh_{x} \over  ch_{x}\mp 1  } 
\begin{pmatrix}
 &  \\[-8pt] - \Im \bm b \, &\Im \tilde {\bm a}\\[-8pt] &  \\
\end{pmatrix} \diag{\sqrt{\frac{M_{22}}{|M|}} }{\sqrt{\frac{1}{M_{22}}}}   .
\end{align}
and it comes down to the solutions~(\ref{Reu}) and (\ref{Imu}).

For the case $|O| = -1$, since $1 \pm O^{T}$ are singular matrices with $\det ( 1\pm O^{T}) = 0$, 
we consider $1 \pm O^{T} \s_{3}$ instead.
This extra $\s_{3}$ exchanges $\Re \bm b$ and $\Im \bm b$. 
For example, in the case of $X^{*} = X O^{T}$, 
\begin{align}
X (1 + O^{T} \s_{3}) & = X + X^{*} \s_{3} = 2 ( \Re \tilde {\bm a} \, , i \, \Im \bm b)  \sqrt{\tilde M ^{-1}} \, ,  \label{xpo} \\
X (1 - O^{T} \s_{3}) &  = X - X^{*} \s_{3} = 2 (i \, \Im \tilde {\bm a} \, , \Re \bm b) \sqrt{\tilde M ^{-1} } \, . \label{xmo}
\end{align}
A similar relationship as Eq.~(\ref{49}) holds for $1 \pm O^{T} \s_{3}$, 
\begin{align}
1 \pm O^{T} \s_{3} & = 1_{2} (1 \pm c_{x})  \mp i \s_{2} s_{x} \, . 
\end{align}
From this, 
\begin{align}
(1 \pm O^{T} \s_{3}) {1 \mp c_{x} \over s_{x}} & = 1_{2} s_{x} \mp i \s_{2} (1 \mp c_{x}) 
= \mp (1 \mp O^{T} \s_{3}) i \s_{2} \, . 
\end{align}
For example, the solution of the upper sign leads to Eqs.~(\ref{Reu2}) and (\ref{Imu2}), 
\begin{align}
( \Re \tilde {\bm a} \, , i \Im \bm b) \sqrt{\tilde M^{-1}} 
= {- s_{x} \over 1 - c_{x}}  (i \Im \tilde {\bm a} \, , \Re \bm b) \sqrt{\tilde M^{-1}}
 \offdiag{1}{-1} 
= {s_{x} \over 1 - c_{x}}  (\sqrt{\frac{1}{M_{22}}} \Re \bm b \, , - i \sqrt{\frac{M_{22}}{|M|}}  \Im \tilde {\bm a}) \, .
\end{align}

The symmetric conditions~(\ref{cond3}) satisfied by these solutions can be displayed in matrices as well. 
When $|O| = -1$, the condition $M_{22}^{2} \, \Re \tilde a_{i} \, \Im \tilde a_{j} = -  |M| \, \Re  b_{i} \, \Im  b_{j}$ is obviously equivalent to $\Re X \Im X^{T} = 0$.
This condition can be rewritten as 
\begin{align}
\Re X \Im X^{T} = {1\over 4i}  X (1\pm O^{T})(1 \mp O^{T})^{T} X^{T} 
=  \pm {1\over 4i}  X (O^{T} - O) X^{T} = 0 \, . 
\end{align}
Since $O^{T} = O$ holds, it is a solution.

On the other hand, when $|O| = +1$, a similar trick as Eqs.~(\ref{xpo}) and (\ref{xmo}) leads to 
a condition 
\begin{align}
{1\over 4 i} X (1 + O^{T} \s_{3}) (1 - O^{T} \s_{3})^{T} X^{T} = 
( \Re \tilde {\bm a} \, , i \Im \bm b) {\tilde M ^{-1}} (i \Im \tilde {\bm a} \, , \Re \bm b)
= 0 \, , 
\end{align}
that is equivalent to $M_{22}^{2} \, \Re \tilde a_{i} \, \Im \tilde a_{j} = -  |M| \, \Im  b_{i} \, \Re  b_{j}$. 
This can be rewritten as
\begin{align}
(1+ O^{T} \s_{3}) (1 - O^{T} \s_{3})^{T} = O^{T} \s_{3} - \s_{3} O = 0 \, .
\end{align}
In this case, it is still a solution because $\s_{3} O^{T} \s_{3} = O$ holds. 

By understanding the solution $X^{*} O = \pm X$ as a product of matrices,  
the form of $\tilde Y$ and $Y$ can be specified. 
First, we show $O_{0} \equiv Q^{\dg} Q$ and $ O_{1} \equiv Q^{\dg} \s_{3} Q$
with a orthogonal matrix $Q$ are Hermitian orthogonal matrices.
Specifically, 
\begin{align}
Q \equiv 
\begin{pmatrix}
\cos w & - \sin w \\
 \sin w & \cos w \\
\end{pmatrix}
~~ \To ~~
O_{0} &= Q^{\dg} Q = 
\begin{pmatrix}
\cos (w - w^{*}) & - \sin (w - w^{*}) \\
\sin (w - w^{*}) & \cos (w - w^{*})
\end{pmatrix} \, , \\
O_{1} &= Q^{\dg} \s_{3} Q = 
\begin{pmatrix}
\cos (w + w^{*}) & - \sin (w + w^{*}) \\
- \sin (w + w^{*}) & - \cos (w + w^{*}) \\
\end{pmatrix} \, .
\end{align}
Indeed $O_{0,1}$ is Hermitian with $\det O_{0,1} = \pm 1$, and they agree with Eq.~(\ref{O}) by suitable redefinitions.
The solution of GCP~(\ref{GCP}) is transformed as follows.
\begin{align}
X^{*} Q^{\dg} = \pm X Q^{T} ~~{\rm or}~~ X^{*} Q^{\dg}  = \pm X Q^{T} \s_{3} \, . 
\end{align}
Then $X Q^{T}$ is one of the four trivial solutions in Eq.~(\ref{trivial}); 
\begin{align}
X Q^{T} = \{ X_{\pm} \, , X_{\pm}' \, \} \, ,  ~~
\{ X_{\pm}^{*} \, , X_{\pm}^{ \prime *} \, \} = \{ \pm X_{\pm} \, , \pm X_{\pm}' \s_3 \, \} \, .
\end{align}
From this $X = X_{\pm}Q$ or $X = X_{\pm}' Q$ holds. 
They certainly satisfy the GCP~(\ref{GCP}); 
\begin{align}
(X_{\pm} Q)^{*} & = \pm X_{\pm} Q^{*} = \pm (X_{\pm} Q) Q^{T} Q^{*} = \pm (X_{\pm} Q) O_{0}^{-1} \, , \\
(X_{\pm}' Q)^{*} & = \pm X_{\pm}' \s_{3} Q^{*} = \pm (X_{\pm}' Q) Q^{T} \s_{3} Q^{*} = \pm (X_{\pm}' Q) O_{1}^{-1}\, . 
\end{align}

Since $X = X_{\pm}^{(\prime)} Q$ holds, the Yukawa matrix is  reconstructed as 
\begin{align}
\tilde Y = X \sqrt{\tilde M} = X_{\pm}^{(\prime)} Q \sqrt{\tilde M} \, , ~~~
Y = \tilde Y L^{-1} = X_{\pm}^{(\prime)} Q \sqrt{\tilde M} L^{-1} \, . 
\end{align}
Recalling that $L^{-1} M^{-1} (L^{T})^{-1} = \tilde M^{-1}$ from Eq.~(\ref{tildeM}), 
we finally obtain 
\begin{align}
m = X_{\pm}^{(\prime)} Q Q^{T} X_{\pm}^{(\prime) T} \, ,
\end{align}
and $m$ is $CP$-invariant.
Although such results may be well known, 
the main results of this paper are the formula~(\ref{formula}) and 
 Eqs.~(\ref{cond3})-(\ref{solution}), the conditions of $CP$-invariance for $m$ in an arbitrary basis.

\section{Summary}

In this paper, defining a formula by $LDL^{T}$ decomposition for the minimal type-I seesaw mechanism,  
we obtain conditions of $CP$-invariance for the neutrino mass matrix $m$ in an arbtirary basis.
The conditions are found to be 
$\Re (M_{22} a_{i} - M_{12} b_{i}) \, \Im ( M_{22} a_{j} - M_{12} b_{j}) = 
- \det M \, \Re  b_{i} \, \Im  b_{j}$ or $ = - \det M \, \Im  b_{i} \, \Re  b_{j}$
for the Yukawa matrix $Y_{ij} = (a_{j}, b_{j})$ and the right-handed neutrino mass matrix $M_{ij}$.
In other words, the real or imaginary part of $b_{i}$ must be proportional to the real or imaginary part of the quantity $(M_{22} a_{i} - M_{12} b_{i})$.

We also discussed the relevance of the generalized $CP$ symmetry that exists in such a solution.
As a result, Yukawa matrix $Y$ found to be restricted to the form $Y = X_{\pm}^{(')} Q \sqrt{\tilde M} L^{-1}$ with $CP$-covariant Yukawa matrices $X_{\pm}^{(')}$, a complex orthogonal matrix $Q$, the diagonal right-handed neutrino mass matrix $\tilde M$ by the $LDL^{T}$ decomposition, and the unitriangular matrix $L$ at that time.
This result can be extended to other seesaw mechanisms, GCPs, and $Z_{2}$ symmetries \cite{Yang:2022yqw}. 

\section*{Acknowledgement}

This study is financially supported 
by JSPS Grants-in-Aid for Scientific Research
No.~18H01210, No. 20K14459,  
and MEXT KAKENHI Grant No.~18H05543.


\begin{thebibliography}{10}

\bibitem{T2K:2021xwb}
T2K collaboration, K.~Abe {\em et~al.},
\newblock Phys. Rev. D {\bf 103}, 112008 (2021), arXiv:2101.03779.

\bibitem{NOvA:2021nfi}
NOvA collaboration, M.~A. Acero {\em et~al.},
\newblock (2021), arXiv:2108.08219.

\bibitem{Minkowski:1977sc}
P.~Minkowski,
\newblock Phys. Lett. {\bf 67B}, 421 (1977).

\bibitem{GellMann:1980v}
M.~Gell-Mann, P.~Ramond, and R.~Slansky,
\newblock Conf. Proc. {\bf C790927}, 315 (1979).

\bibitem{Yanagida:1979as}
T.~Yanagida,
\newblock Conf. Proc. {\bf C7902131}, 95 (1979).

\bibitem{Casas:2001sr}
J.~A. Casas and A.~Ibarra,
\newblock Nucl. Phys. B {\bf 618}, 171 (2001), arXiv:hep-ph/0103065.

\bibitem{Ibarra:2003up}
A.~Ibarra and G.~G. Ross,
\newblock Phys. Lett. B {\bf 591}, 285 (2004), arXiv:hep-ph/0312138.

\bibitem{Barger:2003gt}
V.~Barger, D.~A. Dicus, H.-J. He, and T.-j. Li,
\newblock Phys. Lett. B {\bf 583}, 173 (2004), arXiv:hep-ph/0310278.

\bibitem{Yang:2021arl}
M.~J.~S. Yang,
\newblock will be published in PTEP  (2022), arXiv:2112.14402.

\bibitem{Ma:1998zg}
E.~Ma, D.~P. Roy, and U.~Sarkar,
\newblock Phys. Lett. B {\bf 444}, 391 (1998), arXiv:hep-ph/9810309.

\bibitem{King:1998jw}
S.~F. King,
\newblock Phys. Lett. B {\bf 439}, 350 (1998), arXiv:hep-ph/9806440.

\bibitem{Frampton:2002qc}
P.~H. Frampton, S.~L. Glashow, and T.~Yanagida,
\newblock Phys. Lett. B {\bf 548}, 119 (2002), arXiv:hep-ph/0208157.

\bibitem{Xing:2020ald}
Z.-z. Xing and Z.-h. Zhao,
\newblock Rept. Prog. Phys. {\bf 84}, 066201 (2021), arXiv:2008.12090.

\bibitem{Guo:2003cc}
W.-l. Guo and Z.-z. Xing,
\newblock Phys. Lett. B {\bf 583}, 163 (2004), arXiv:hep-ph/0310326.

\bibitem{Mei:2003gn}
J.-w. Mei and Z.-z. Xing,
\newblock Phys. Rev. D {\bf 69}, 073003 (2004), arXiv:hep-ph/0312167.

\bibitem{Chang:2004wy}
S.~Chang, S.~K. Kang, and K.~Siyeon,
\newblock Phys. Lett. B {\bf 597}, 78 (2004), arXiv:hep-ph/0404187.

\bibitem{Guo:2006qa}
W.-l. Guo, Z.-z. Xing, and S.~Zhou,
\newblock Int. J. Mod. Phys. E {\bf 16}, 1 (2007), arXiv:hep-ph/0612033.

\bibitem{Kitabayashi:2007bs}
T.~Kitabayashi,
\newblock Phys. Rev. D {\bf 76}, 033002 (2007), arXiv:hep-ph/0703303.

\bibitem{He:2009pt}
X.-G. He and W.~Liao,
\newblock Phys. Lett. B {\bf 681}, 253 (2009), arXiv:0909.1463.

\bibitem{Yang:2011fh}
R.-Z. Yang and H.~Zhang,
\newblock Phys. Lett. B {\bf 700}, 316 (2011), arXiv:1104.0380.

\bibitem{Harigaya:2012bw}
K.~Harigaya, M.~Ibe, and T.~T. Yanagida,
\newblock Phys. Rev. D {\bf 86}, 013002 (2012), arXiv:1205.2198.

\bibitem{Kitabayashi:2016zec}
T.~Kitabayashi and M.~Yasu\`e,
\newblock Phys. Rev. D {\bf 94}, 075020 (2016), arXiv:1605.04402.

\bibitem{Bambhaniya:2016rbb}
G.~Bambhaniya, P.~S. Bhupal~Dev, S.~Goswami, S.~Khan, and W.~Rodejohann,
\newblock Phys. Rev. D {\bf 95}, 095016 (2017), arXiv:1611.03827.

\bibitem{Li:2017zmk}
C.-C. Li and G.-J. Ding,
\newblock Phys. Rev. D {\bf 96}, 075005 (2017), arXiv:1701.08508.

\bibitem{Liu:2017frs}
Z.-C. Liu, C.-X. Yue, and Z.-h. Zhao,
\newblock JHEP {\bf 10}, 102 (2017), arXiv:1707.05535.

\bibitem{Shimizu:2017fgu}
Y.~Shimizu, K.~Takagi, and M.~Tanimoto,
\newblock JHEP {\bf 11}, 201 (2017), arXiv:1709.02136.

\bibitem{Shimizu:2017vwi}
Y.~Shimizu, K.~Takagi, and M.~Tanimoto,
\newblock Phys. Lett. B {\bf 778}, 6 (2018), arXiv:1711.03863.

\bibitem{Nath:2018hjx}
N.~Nath, Z.-z. Xing, and J.~Zhang,
\newblock Eur. Phys. J. {\bf C78}, 289 (2018), arXiv:1801.09931.

\bibitem{Barreiros:2018bju}
D.~M. Barreiros, R.~G. Felipe, and F.~R. Joaquim,
\newblock JHEP {\bf 01}, 223 (2019), arXiv:1810.05454.

\bibitem{Nath:2018xih}
N.~Nath,
\newblock Mod. Phys. Lett. A {\bf 34}, 1950329 (2019), arXiv:1808.05062.

\bibitem{Wang:2019ovr}
X.~Wang and S.~Zhou,
\newblock JHEP {\bf 05}, 017 (2020), arXiv:1910.09473.

\bibitem{Zhao:2020bzx}
Z.-h. Zhao,
\newblock JHEP {\bf 11}, 170 (2021), arXiv:2003.00654.

\bibitem{Ecker:1981wv}
G.~Ecker, W.~Grimus, and W.~Konetschny,
\newblock Nucl. Phys. B {\bf 191}, 465 (1981).

\bibitem{Ecker:1983hz}
G.~Ecker, W.~Grimus, and H.~Neufeld,
\newblock Nucl. Phys. B {\bf 247}, 70 (1984).

\bibitem{Gronau:1985sp}
M.~Gronau and R.~N. Mohapatra,
\newblock Phys. Lett. B {\bf 168}, 248 (1986).

\bibitem{Ecker:1987qp}
G.~Ecker, W.~Grimus, and H.~Neufeld,
\newblock J. Phys. A {\bf 20}, L807 (1987).

\bibitem{Neufeld:1987wa}
H.~Neufeld, W.~Grimus, and G.~Ecker,
\newblock Int. J. Mod. Phys. A {\bf 3}, 603 (1988).

\bibitem{Ferreira:2009wh}
P.~Ferreira, H.~E. Haber, and J.~P. Silva,
\newblock Phys. Rev. D {\bf 79}, 116004 (2009), arXiv:0902.1537.

\bibitem{Feruglio:2012cw}
F.~Feruglio, C.~Hagedorn, and R.~Ziegler,
\newblock JHEP {\bf 07}, 027 (2013), arXiv:1211.5560.

\bibitem{Holthausen:2012dk}
M.~Holthausen, M.~Lindner, and M.~A. Schmidt,
\newblock JHEP {\bf 04}, 122 (2013), arXiv:1211.6953.

\bibitem{Ding:2013bpa}
G.-J. Ding, S.~F. King, and A.~J. Stuart,
\newblock JHEP {\bf 12}, 006 (2013), arXiv:1307.4212.

\bibitem{Girardi:2013sza}
I.~Girardi, A.~Meroni, S.~Petcov, and M.~Spinrath,
\newblock JHEP {\bf 02}, 050 (2014), arXiv:1312.1966.

\bibitem{Nishi:2013jqa}
C.~Nishi,
\newblock Phys. Rev. D {\bf 88}, 033010 (2013), arXiv:1306.0877.

\bibitem{Ding:2013hpa}
G.-J. Ding, S.~F. King, C.~Luhn, and A.~J. Stuart,
\newblock JHEP {\bf 05}, 084 (2013), arXiv:1303.6180.

\bibitem{Feruglio:2013hia}
F.~Feruglio, C.~Hagedorn, and R.~Ziegler,
\newblock Eur. Phys. J. C {\bf 74}, 2753 (2014), arXiv:1303.7178.

\bibitem{Chen:2014wxa}
P.~Chen, C.-C. Li, and G.-J. Ding,
\newblock Phys. Rev. D {\bf 91}, 033003 (2015), arXiv:1412.8352.

\bibitem{Ding:2014ora}
G.-J. Ding, S.~F. King, and T.~Neder,
\newblock JHEP {\bf 12}, 007 (2014), arXiv:1409.8005.

\bibitem{Ding:2014hva}
G.-J. Ding and Y.-L. Zhou,
\newblock JHEP {\bf 06}, 023 (2014), arXiv:1404.0592.

\bibitem{Chen:2014tpa}
M.-C. Chen, M.~Fallbacher, K.~Mahanthappa, M.~Ratz, and A.~Trautner,
\newblock Nucl. Phys. B {\bf 883}, 267 (2014), arXiv:1402.0507.

\bibitem{Chen:2015siy}
P.~Chen, G.-J. Ding, F.~Gonzalez-Canales, and J.~W.~F. Valle,
\newblock Phys. Lett. {\bf B753}, 644 (2016), arXiv:1512.01551.

\bibitem{Li:2015jxa}
C.-C. Li and G.-J. Ding,
\newblock JHEP {\bf 05}, 100 (2015), arXiv:1503.03711.

\bibitem{Turner:2015uta}
J.~Turner,
\newblock Phys. Rev. D {\bf 92}, 116007 (2015), arXiv:1507.06224.

\bibitem{Rodejohann:2017lre}
W.~Rodejohann and X.-J. Xu,
\newblock Phys. Rev. {\bf D96}, 055039 (2017), arXiv:1705.02027.

\bibitem{Penedo:2017vtf}
J.~Penedo, S.~Petcov, and A.~Titov,
\newblock JHEP {\bf 12}, 022 (2017), arXiv:1705.00309.

\bibitem{Nath:2018fvw}
N.~Nath, R.~Srivastava, and J.~W. Valle,
\newblock Phys. Rev. D {\bf 99}, 075005 (2019), arXiv:1811.07040.

\bibitem{Yang:2020qsa}
M.~J.~S. Yang,
\newblock Phys. Lett. B {\bf 806}, 135483 (2020), arXiv:2002.09152.

\bibitem{Yang:2020goc}
M.~J.~S. Yang,
\newblock Chin. Phys. C {\bf 45}, 043103 (2021), arXiv:2003.11701.

\bibitem{Yang:2021smh}
M.~J.~S. Yang,
\newblock Nucl. Phys. B {\bf 972}, 115549 (2021), arXiv:2103.12289.

\bibitem{Yang:2021xob}
M.~J.~S. Yang,
\newblock PTEP {\bf 2022}, 013 (2021), arXiv:2104.12063.

\bibitem{Yang:2021byq}
M.~J.~S. Yang,
\newblock will be published in PTEP  (2022), arXiv:2110.10907.

\bibitem{Chen:2016ptr}
P.~Chen, G.-J. Ding, and S.~F. King,
\newblock JHEP {\bf 03}, 206 (2016), arXiv:1602.03873.

\bibitem{Zhao:2021dwc}
Z.-h. Zhao,
\newblock (2021), arXiv:2111.12639.

\bibitem{Yang:2022yqw}
M.~J.~S. Yang,
\newblock (2022), arXiv:2204.08607.

\end{thebibliography}

\end{document}